\documentclass[letterpaper,12pt]{article}
\usepackage[USenglish]{babel} 
\usepackage{setspace}
\usepackage{lmodern} 
\usepackage{endnotes}
\usepackage[round]{natbib}

\usepackage{graphicx} 
\usepackage{amsmath}
\usepackage{amsthm}
\usepackage{amsfonts}

\let\footnote=\endnote
\begin{document}

\pagestyle{empty} 

\title{A theoretical framework for conducting multi-level studies of complex social systems with agent-based models and empirical data}
\author{Chih-Chun Chen}
\date{}
\maketitle
\newpage
\begin{abstract}
A formal but intuitive framework is introduced to bridge the gap between data obtained from empirical studies and that generated by agent-based models. This is based on three key tenets. Firstly, a simulation can be given multiple formal descriptions corresponding to static and dynamic properties at different levels of observation. These can be easily mapped to empirically observed phenomena and data obtained from them. Secondly, an agent-based model generates a set of closed systems, and computational simulation is the means by which we sample from this set. Thirdly, properties at different levels and statistical relationships between them can be used to classify simulations as those that instantiate a more sophisticated set of constraints. These can be validated with models obtained from statistical models of empirical data (for example, structural equation or multi-level models) and hence provide more stringent criteria for validating the agent-based model itself. 


\end{abstract}

\cleardoublepage 

\pagestyle{myheadings}
\markboth{ABM framework for multi-level studies of complex social systems}{ABM framework for multi-level studies of complex social systems}

\section{Introduction}
\label{sec:intro}
Many social and economic phenomena can be characterised in terms of `complex systems'. Within this characterisation, entities and patterns of behaviour emerge at different levels, and interact with one another in non-linear ways. Agent-based modelling (ABM) is a computational method for modelling and simulating such systems. 

The complex systems perspective has two major strands. From a more Statistical Mechanics-oriented view, the study of complex systems has focused mainly on the ways in which lower level micro-properties and interactions give rise to higher level macro-properties \citep{feldman97}, \citep{ellis05}. The more biologically-based approach tends to focus more on relating properties at different levels, such as functional modules in the brain or biochemical pathways and networks (in some cases, such as feedback, the emergent phenomenon may even be at the micro-level) \citep{varela79}, \citep{tononi94}, \citep{hartwell99}. Both these strands should be leveraged in the social sciences. 

From a policy point of view, it is important to understand how changes in rules at the micro-level (which might represent the interaction between psychology and policy) affect more macro-level behaviours (which might include those associated with family, organisational, or geographical units). At the same time, we often have important information about the way decisions or behaviours of units at different levels relate to each other (for example, if several commercial organisations dominate a sector, this can affect both other organisations within the sector and other sectors).  

Currently, the complex systems approach is largely model-dominated and/or model-driven (whether models are informal, formal, mathematical, or statistical). While this allows theories to be specified with precision, there is a risk of alienating those conducting empirical research and hence models becoming irrelevant or too idealized for real world application. There is therefore an urgent need to establish robust techniques for analysing and validating models with respect to empirical data, particularly as, unlike in the physical sciences, idealizations of models do not always have clear isomorphism with empirically based studies \citep{henrickson02}. As ABM is maturing and becoming more widely adopted in the social sciences \citep{bonabeau02}, \citep{sawyer01}, \citep{gilbert05}, \citep{economist10}, it is crucial that the appropriate methods of analysis are applied and that the conclusions we draw from these analyses are valid. This requires an understanding of their theoretical basis and rationale. 

Furthermore, a rigorously grounded theory allows us to defend the conclusions we draw from validating models against empirical data and avoid the doubts often cast upon the utility and validity of agent-based models (see, for example \citep{mccauley07}). Related to this are questions regarding the interpretation and analysis of \textit{simulations}, for example:

\begin{itemize}
\item How many simulations do we need to run to draw a conclusion?
\item How do we interpret differences between simulations?
\item How do we use empirical data and simulation-generated data to validate a model? (Discussions of different aspects of ABM empirical validation issues can be found in \citep{kleijnen90},, \citep{axtell96a}, \citep{kleijnen01}, \citep{fagiolo03}, \citep{troitzsch04}, \citep{brenner07}, \citep{marks07}, \citep{windrum07}, \citep{moss08}.) 
\item How do we choose between different agent-based models and parameter configurations when they are all able to generate empirically valid data? 
\end{itemize}

This article introduces a simple theory of types for describing agent-based simulations at different levels and relates this to the application of different established analytical techniques. The theory is based on three fundamental tenets:

\begin{enumerate}
\item Theoretically, an agent-based model generates a finite set of formally describable closed complex systems, and simulations are the means by which we sample from this set. In other words, each simulation is an instantiation of a possible system generated by the model;
\item A simulation can be formally described in terms of properties or phenomena at different levels, with micro-level properties corresponding to computational states end events, and higher level properties corresponding to sets and/or structures of these states and events. (Higher level properties such as population behaviour can also be expressed in terms of macro-variables and changes in macro-variables, which would define the sets of event structures; see Section \ref{subsec:eqToCET})
\item Property descriptions at different levels can be used (either in isolation or in combination) to classify simulations.
\end{enumerate}
 
These tenets allow us to build more stringently constrained models relating phenomena at different levels, which provide stricter criteria for validation with empirical data. Instead of simply requiring that some phenomenon `emerges' at the systemic level in simulations, \textit{structures} of \textit{related phenomena} (possibly at different scales and/or levels of abstraction) need to be reproduced with \textit{appropriate frequencies or probabilities}.

Before commencing, we wish to emphasize that the social sciences cover a vast landscape of disciplines and domains, and that each domain (and subdomain) will have its own set of issues to address when both developing and validating agent-based models. The hope is that each specific domain will be able to adapt, extend and apply our framework for their specific purposes.

\section{Background and motivation: The application of ABM in the study of social systems}

Quantitative characterization of dynamic social and economic systems is often problematic because such systems are complex. By complex, it is meant that the behaviour of these systems arises as the result of interactions between multiple factors at different levels (we will formalise the notion of levels in Section \ref{sec:observation}). In the Complex Systems literature (particularly from Statistical Physics) the terms `non-equilibrium', `non-linear' and `non-ergodic' are often used to refer to describe such system. The difficulty posed by such systems is that knowledge of micro-behaviour does not guarantee knowledge of the macro-behaviour, and vice versa. There are two aspects to this. 

Firstly, the macro-level behaviour  by definition can not be descriptively or logically reduced to micro-level behaviour; language used to describe the micro-level is therefore logically distinct from that used to characterise the macro-level \citep{darley94}, \citep{bonabeau97}, \citep{kubik03}, \citep{deguet06}. This follows from the fact that micro- and macro-level phenomena require different levels of observation to be manifest \citep{crutchfield94}, \citep{crutchfield03}, \citep{sasai08}, \citep{ryan07}, \citep{prokopenko08}. 

Secondly, in contrast to systems in equilibrium in which differences at the micro-level make little difference to macro-level observations and hence for which we can predict macro-level behaviour from micro-level observations, complex systems are sensitive to relatively small perturbations. This sensitivity means that perturbations at the micro-level can have non-linear effects at the macro-level \citep{kauffman93}, \citep{holland00},  \citep{baryam03}, \citep{ellis05}. 

The motivation for ABM comes from both these aspects. In ABM, the micro-level is specified computationally in the form of state transition rules ($STR$s) governing the behaviour of computational agents and the macro-level behaviour is usually represented by system-level global state variables which aggregate in some way the states or behaviours of the agents in the system. These two modes of representation can be seen to respectively represent the logically distinct micro- and macro-level languages. At the same time, ABM is used to study the effects of perturbations at the micro-level, which are introduced as differences in initial conditions and/or parameters. The types of questions that ABM practice typically try to address are: 
\begin{itemize}
	\item How different are the behaviours of simulations generated from different initial conditions?
	\item Which parameters is the model most sensitive to?
	\item Under which parameter configurations and/or ranges is the behaviour most sensitive or stable? 
\end{itemize}

However, the idea that complex, non-linear relationships exist between phenomena at different levels is in fact extremely pervasive in empirical studies. The key difference between such studies and more model-centric approaches to complexity lies in the methods used to analyse and represent this complexity. In empirical studies, the techniques tend to focus more on interactive statistical associations between phenomena e.g. \citep{pearl98}, \citep{krull01}, which tend to be represented in network-based or hierarchical models, such as Bayesian networks, structural equations, or multi-level models. In such representations, the relations between `levels' are not formal, but descriptive (based only on our understanding of the phenomena) or statistical (as in the case of multi-level \citep{gelman06}, \citep{gelman06a} or modular \citep{seth08} models). Model-driven studies on the other hand, tend to consider associations in terms of their fundamental statistical mechanics \citep{prokopenko09} or emergent network dynamics \citep{barabasi02}, \citep{dorogovstev03}. 

This paper seeks to explicitly relate these two perspectives using an extended ABM framework that permits the representation of properties and behaviours at any level of abstraction and the relationships between them, going beyond simple two-level micro-macro/macro-micro relationships.

\subsection{Hypotheses, empirical data, models and simulations}
\label{sec:modelSimulations}

\begin{figure}
	\centering
		\includegraphics[width=.4\columnwidth]{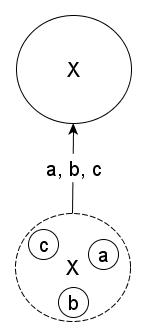}
	\caption{Graphical representation of hypothesis that mechanisms and/or interactions $a$, $b$ and $c$ at the micro-level give rise to phenomenon $X$ at the systemic macro-level.}
	\label{fig:hyp1}
\end{figure}

\begin{figure}
	\centering
		\includegraphics[width=.8\columnwidth]{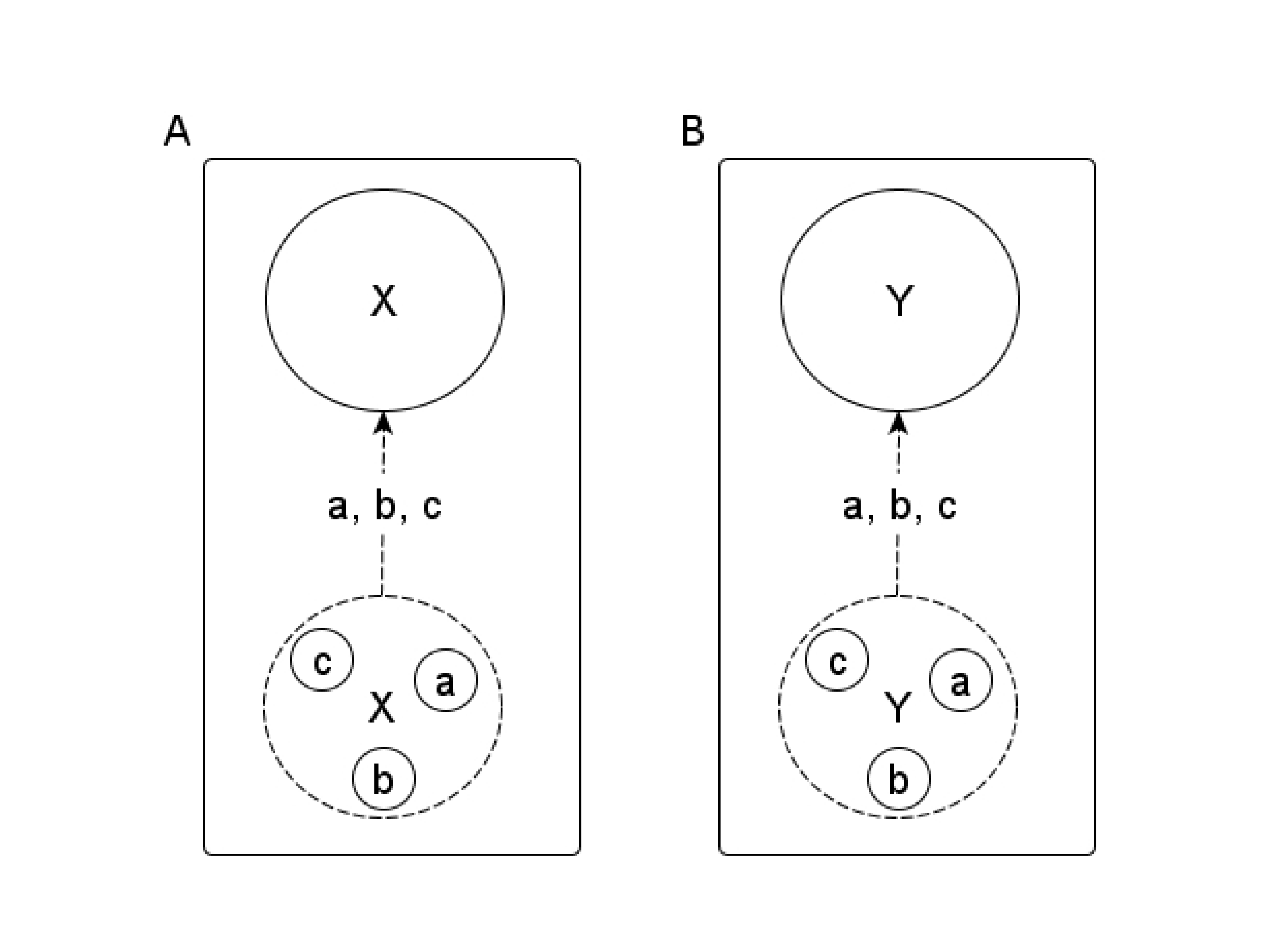}
	\caption{Graphical representation of hypothesis that under condition $A$, mechanisms and/or interactions $a$, $b$ and $c$ at the micro-level give rise to phenomenon $X$ at the systemic macro-level, but under condition $B$, $a$, $b$ and $c$ give rise to phenomenon $Y$.}
	\label{fig:hyp2}
\end{figure}

Empirical validation is a significant challenge that needs to be overcome in order for ABM to become more seriously adopted in the Social sciences. We can classify validation techniques according to the types of hypotheses they support. To date, the motivation for applying agent-based modelling tends to be motivated by the following two hypotheses classes:

\begin{itemize}
\item Hypotheses concerning the ability of mechanisms and interactions at the micro-level to give rise to phenomena at the systemic level, for example, attraction/repulsion -> regional segregation. In these cases, qualitative data can be used to (weakly) validate the model (i.e. show that it is not false). At the micro-level, these might be based on findings from Psychology or on enforced policies. At the systemic level, they might be anecdotal or event-based observations. This is illustrated in Figure \ref{fig:hyp1}.
\item Hypotheses concerning the conditions under which mechanisms and interactions at the micro-level are able to give rise to phenomena at the systemic level, for example, attraction/repulsion -> regional segregation when the initial diversity of agents exceeds a particular threshold. With these types of hypotheses, validation would require empirical data about both the initial configuration and the observed phenomena (e.g. regional distribution of different ethnic backgrounds at $t_{1}$ and $t_{2}$). This is illustrated in Figure \ref{fig:hyp2}.
\end{itemize}

\begin{figure}
	\centering
		\includegraphics[width=.8\columnwidth]{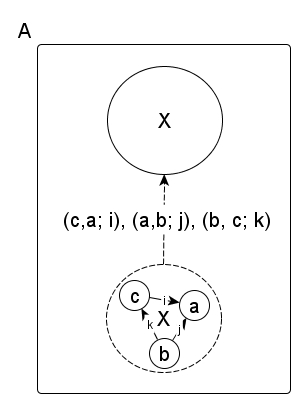}
	\caption{Graphical representation of hypothesis that mechanisms and/or interactions $a$, $b$ and $c$ at the micro-level need to be related by specific associations, represented by $i$, $j$ and $k$, to give rise to phenomenon $X$ at the systemic macro-level. }
	\label{fig:hyp3}
\end{figure}

\begin{figure}
	\centering
		\includegraphics[width=.8\columnwidth]{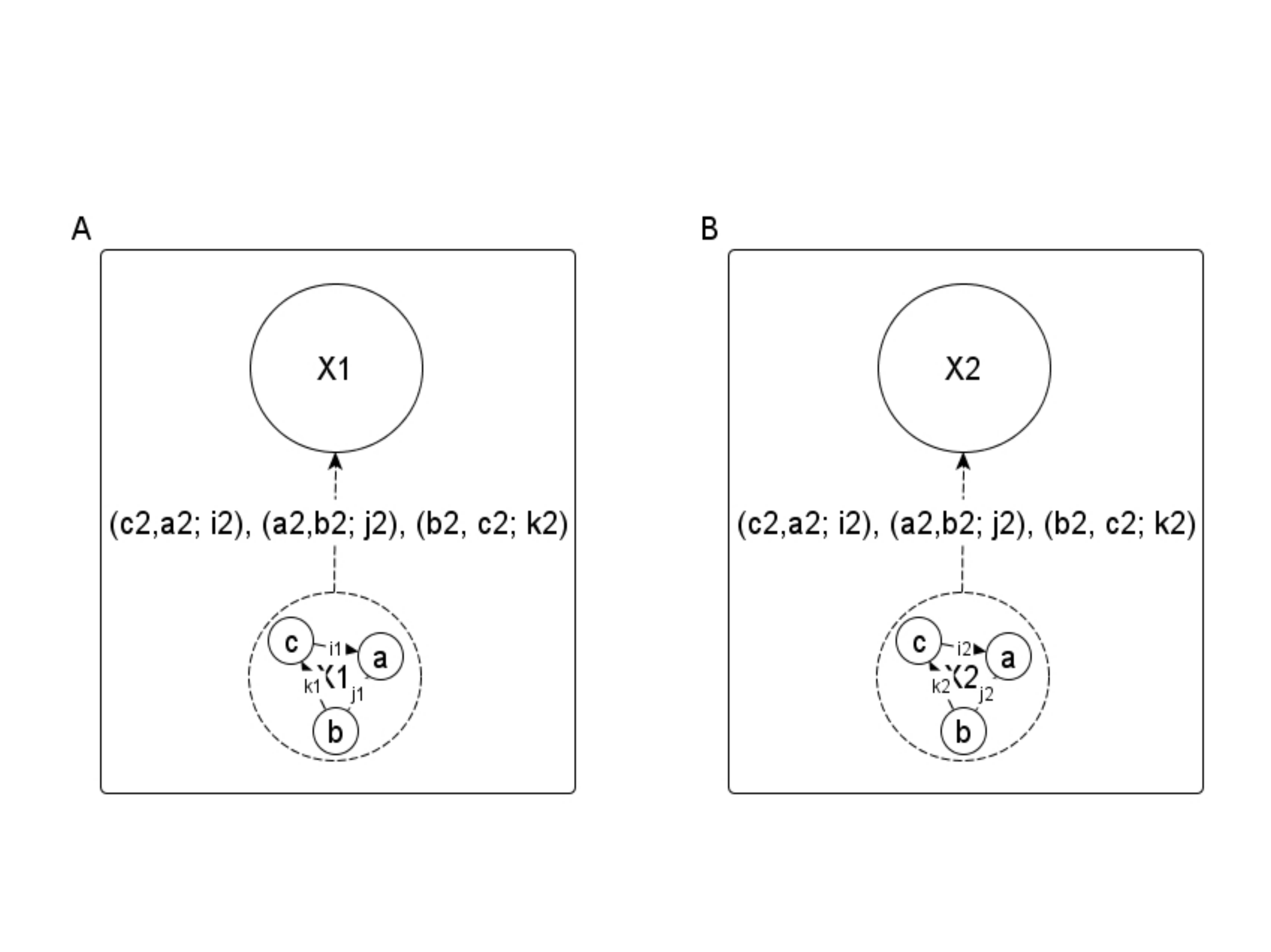}
	\caption{Graphical representation of hypothesis that under condition $A$, mechanisms and/or interactions $a$, $b$ and $c$ at the micro-level need to be related by specific associations, represented by $i1$, $j1$ and $k1$, to give rise to phenomenon $X$ at the systemic macro-level, but under condition $B$, they need different relations $i2$, $j2$, and $k2$. This is a multi-level model, where each of the sub-models is distinguished only by the strengths of the relationships (and not the structure).}
	\label{fig:hyp3M}
\end{figure}

\begin{figure}
	\centering
		\includegraphics[width=.8\columnwidth]{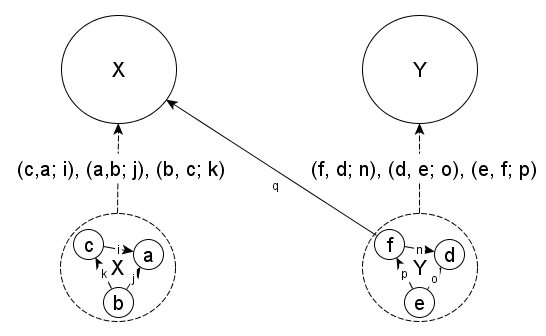}
	\caption{Graphical representation of hypothesis that (i) mechanisms and/or interactions $a$, $b$ and $c$ at the micro-level need to be related by specific associations, represented by $i$, $j$ and $k$, to give rise to phenomenon $X$ at the systemic macro-level; (ii) mechanisms and/or interactions $d$, $e$ and $f$ at the micro-level need to be related by specific associations, $n$, $o$, $p$, to give rise to phenomenon $Y$; and (iii) $Y$ is associated with $X$ by relation $q$. $X$ and $Y$ could also represent phenomena at different abstraction levels}
	\label{fig:hyp4}
\end{figure}

However, we can also analyse agent-based models with empirical data to address the following:
\begin{itemize}
	\item Hypotheses relating micro-level mechanisms to \textit{relationships} between phenomena at \textit{different} levels, including how they might interact to give rise to global systemic phenomena. For example, we could formulate and validate a model that describes the relationship between individuals' psychology, policy decisions, regional migration, local unemployment, and the country's economy. This would require empirical data relating to each of the phenomena at the different levels. If the totality of qualitative effects are observed, then we can say the model is valid. This can be expressed as a graph or network. If the data we have is quantitative, the edges of the graph can also be weighted to represent the strength of the relationships. This is illustrated in Figures \ref{fig:hyp3} and \ref{fig:hyp4}.
	\item Hypotheses about the conditions under which relationships between phenomena at different levels hold. This would require data from different instances of the related phenomena, including their non-occurrence. For example, we would need to ensure that the cases in which the relationships hold have the same features (or feature combinations) as hypothesised, and that these features combinations are not found in the cases in which the relationships do not hold. It is also possible that this a matter of degree e.g. factor $X$ reduces the strength of association between phenomena $A$, $B$, and $C$. This is illustrated in Figure \ref{fig:hyp3M}.
\end{itemize}

\subsection{A general characterisation of ABM}
\label{subsec:abmGeneral}
To ensure we have as general a characterisation of ABM as possible, we do not base our framework on any specific modelling language or software framework, but instead give an abstract definition that can be easily mapped to existing ABM frameworks. 

We define an ABM as a set of agent types $A_{0},..., A_{n}$ (global state variables, e.g. representing resource availability, and dynamic spatial representations can also be represented as agent types in this abstract formulation) and constraints $C$ determining how agents are able to interact in the system (for example, whether they can communicate directly, synchronously, asynchronously, symmetrically, asymmetrically or via some specified protocol or topology; this also determines the updating or execution order). Each agent type $A_{i}$ consists of a set of variables with defined value ranges and a set of state transition rules ($STR_{i}$). $STR$s can be seen to represent the range of possible behaviours for agents (instantiations) of the particular type $A_{i}$ and therefore encode the knowledge we have about individual- or micro-level behaviour. The set of variables and value ranges define the set of states that agents of the type are able to realise.   

We define a state transition rule $STR_{Ai}$ to be a function that maps (i) a source subsystem state ($\varphi_{source}$) represented by the values of some subset of the system's state variables (which might be encapsulated in the agent itself or belong to other agents and/or elements in the system) to (ii) a target subsystem state ($\varphi_{target}$) represented by some new set of values for the set of variables when a particular condition $cn$ is satisfied. The mapping $\varphi_{source}\rightarrow\varphi_{target}$ is the state transition, as defined below:

\paragraph{State transition} A state transition is a transformation of one subsystem state to another subsystem state. The state before the transformation is applied is called the source state and is denoted $\varphi_{source}$, while the state after the transformation has been applied is called the target state and  denoted $\varphi_{target}$. (The definition for subsystem state is given in Definition \ref{def:subsystemState}).

\paragraph{State transition rule ($STR$)} \begin{equation}
STR_{Ai}(cn) = \varphi_{source} \rightarrow \varphi_{target},
\end{equation}
where $cn \in CN$, and $CN$ denotes the set of conditions that can be distinguished by agents of the type $A_{i}$.

\paragraph{*}The condition $cn$ under which an $STR$ is executed might be dependent on the agent's own state $q_{a}$, the state $q_{e}$ of its environment or neighbourhood $e$ (which might itself be made up of other agents' states), or both. State transition rules might also be expressed implicitly in terms of constraints on permissible action as well as explicitly in terms of conditional state changes, but these are formally equivalent.

In the most general terms therefore (abstracting away from particular formal languages or modelling frameworks), an agent-based model is a set of agent types with a set of constraints governing the interactions between agents.

\section{Agent-based models as both generators and classifiers of system types}
\label{sec:abmCSGenerator}
To truly understand what we are doing when we run simulations of agent-based models, it is necessary to delve a little into some of the technical computational details of simulation. Although the practice of agent-based modelling should be seen as abstracted from computational matters (just as programming languages are seen as distinct from machine code), when running simulations, the realisation of computations can have certain implications.

The following are especially important to note:

\paragraph{Execution order}
Different orders of execution and state updating can lead to radically different outcomes, even with the same initial conditions and parameter settings \citep{gargi08}, \citep{blok99}. In fact, we can see different updating rules as an extension of the agent-based model itself, since the set of systems generated by one set of updating rules (e.g. asynchronous) is different to (and may not even overlap with) the set generated by another (e.g. synchronous).

\paragraph{Set of systems generated}The set of possible systems (distinguishable simulations) that can be generated from an agent-based model can be arbitrarily limited by the nature of the platform on which it is run. This is particularly pertinent in cases where real (rather than integer) values are included in the model or where stochasticity features. In the case of real values, the memory limitations mean that accuracy is limited. In other words, the set of possible simulations only includes systems in which we are able to measure a variable to $n$ decimal places. While this might at first seem trivial, the implication is theoretically significant, since it means that the set of systems we are able to study computationally (if we were to simulate every possible system) is only a subset of the possible systems that could theoretically be generated by our agent-based model.

\paragraph*{}More generally, if we see each distinguishable simulation \footnote{Two simulation instances with the same sequence and structure of agent rule executions and resulting state changes are indistinguishable.} as a computational representation of a system that the agent-based model can generate, it is clear that however many simulations we run, the set of systems that we can study is finite, even if the agent-based model is theoretically able to generate an infinite set of systems. (We will formalise this later in terms of complex event types.) \footnote{It is important to note that in many cases, the agent-based model itself implies a finite set of systems e.g. a closed system with boolean values deterministically governing rule execution.} 

In the remainder of this section we will probe more deeply into the implications of this for three important aspects of simulation: (i) model concretisation for validating predicted behaviour; (ii) sampling to determine `typical' behaviour; (iii) probing to evaluate parameter sensitivity.

\subsection{Simulation as model concretization}
\label{subsec:concretizingSimulation}

In the practice of agent-based modelling, the most basic function of simulation is to establish whether or not the model defined at the agent level is able to generate some phenomenon at a higher systemic level. This is typically represented by one or more state variables that aggregate individual agents' state variable values. In many cases, models are also parametised so as to capture some features of the system being modelled, so that the higher level phenomenon is hypothesised to occur within some defined value range(s). Simulation is therefore treated as a means of determining what happens when the statically represented agent-based model (expressed in terms of agent state transition rules) is concretised and dynamically executed under particular conditions (represented by parameters and initial conditions).  

Returning to the fact that the set of distinguishable simulations is finite, the implication is that even if we were to run every possible simulation, we never observe the desired systemic phenomenon even though the agent-based model is theoretically able to generate it. In other words, we are only able to concretise part of the agent-based model (this is equivalent to saying we can only sample a subset of the possible systems the model can generate; see below).     

This is especially problematic when the phenomenon we are trying to understand itself a one-off or rare event. In this case, we have no information about how probable the phenomenon is under the conditions we have represented in the concretised model. Hence, even if the concretised model (simulation) does emulate the phenomenon, we are not really entitled to draw any strong conclusions (unless we have extremely detailed information about the initial conditions and the phenomenon is only reproduced in simulations where these initial conditions are realised; this is the rationale behind `history-friendly' validation \citep{werker04}).

\subsection{Simulation as sampling}
\label{subsec:samplingSimulation}
\begin{figure}
	\centering
		\includegraphics[width=.9\columnwidth]{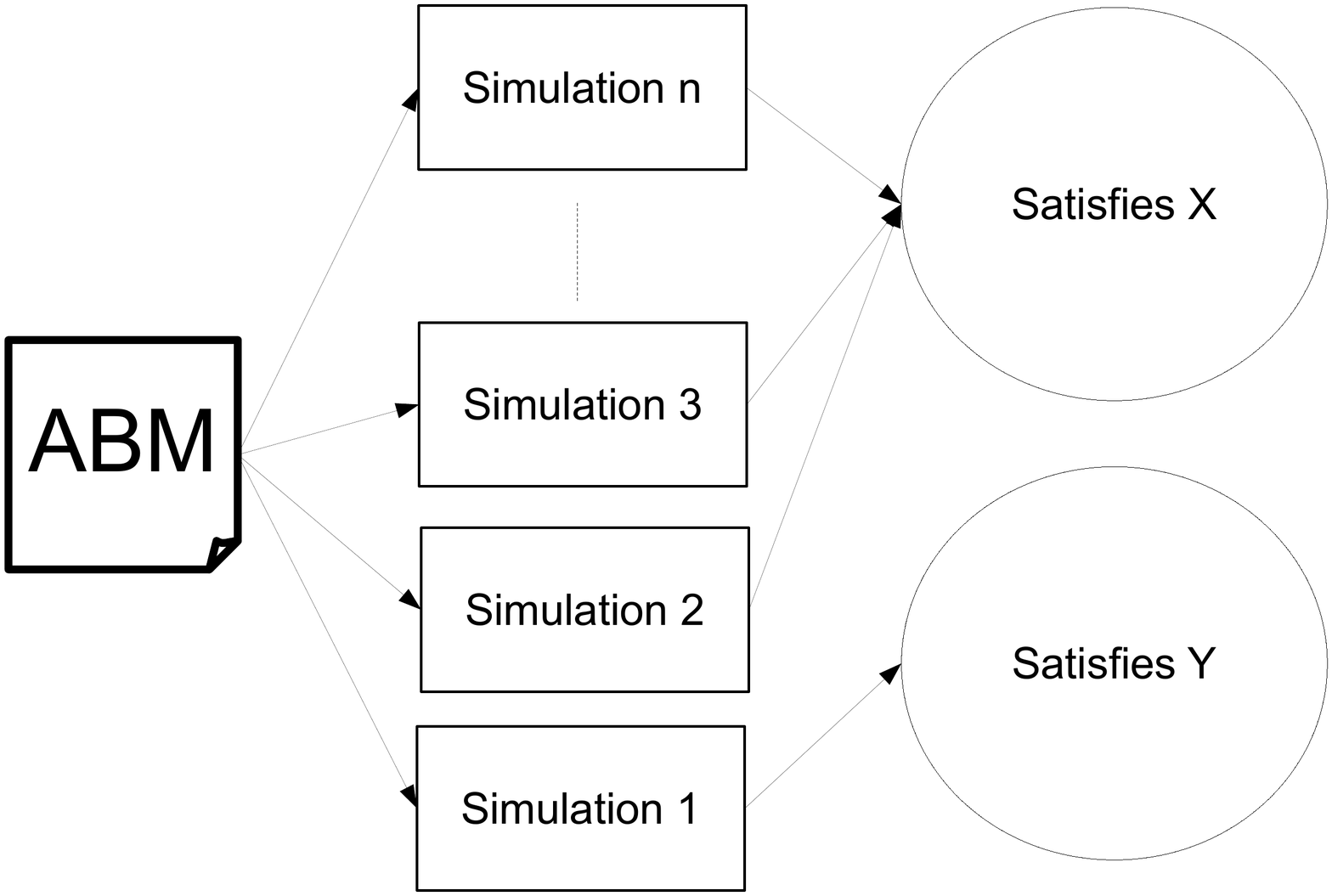}
	\caption{An agent-based model (abm) generates a set of possible system trajectories, of which a simulation is an instantiation. The occurrence rate of simulations with a particular set of attributes (X and Y) reflects the probability or frequency with which this type of system is expected to occur given the agent-based model. Attributes X and Y could include any combination of within-simulation observations and measures discussed above in Section \ref{sec:observation}, such as the the emergence of a particular global phenomenon or end state.}
	\label{fig:popSims}
\end{figure}

Another widely adopted approach to simulation is to treat it as sampling (see Figure \ref{fig:popSims}). In terms of data about the system being modelled, this requires us to have information about the distribution or probability with which the desired phenomenon occurs. When simulating therefore, it is not sufficient simply to reproduce the phenomenon, but to reproduce it to the correct degree. For example, if our real world data tell us that phenomenon $X$ occurs in $50\%$ of the cases, only around $50\%$ of our simulations should exhibit the  phenomenon (assuming that we have represented in our agent-based model everything we know about the system and that the fact that $X$ is only observed in $50\%$ of the empircal cases is due to the incompleteness of our knowledge of the conditions necessary for it to occur).

The issue with sampling from only a subset of systems implied by the agent-based model is that neither our knowledge nor our ignorance is completely represented. Hence the resulting distribution of simulations sampled is not strictly speaking a reflection of the information (or lack of information) we have included in the agent-based model.

\subsection{Simulation as probing}
\label{subsec:probingSimulation}
Yet another approach to agent-based simulation is to use it as a means of understanding the fundamental nature of the phenomenon being studied. This is strongly linked to other complex systems modelling techniques, such as equations or iterative maps. The type of model features that we are interested in within this approach include for example, whether or not a phenomenon is sensitive to scale (scale invariance) or how the degree to which it occurs alters under different conditions (parameter sensitivity). In other words, simulation is used as a means to better understand the \textit{model} and its set of systems. 

The issue that arises here generalises that which arises when simulation is used as a means of sampling. If we are using simulation as a means of understanding the shape of the space of systems defined by our model, the fact that we may only able to include a subset of the possible systems means that only a region of the possible locations in the space of systems will be accessible to us, leading to a mis-representation of the shape of this space.    

\paragraph*{}More concretely, our response to the result that out of 1000 simulations, all expect one show sufficient agreement with our empirical data might be very different depending on the type of study. We could conclude that we have captured the essential mechanisms underlying the phenomenon described by our empirical data and that our agent-based model has been validated. On the other hand, we might wish to further investigate the differences between the anomalous simulation and the others by identifying the key differences (for example, different initial conditions, subsystem behaviours or global sub-trajectories). Empirical data associated with these distinguishing attributes could then be sought to provide further support for the model (in the best case, the differences in the anomalous simulation would map directly onto an anomalous case in the real world with the same distinguishing attributes). 

On the other hand, even if the distinguishing attributes in the anomalous simulation are implausible (for example, they refute what we believe should be possible in human interactions), we might still accept the model as having sufficient explanatory and predictive validity since the vast majority of simulations manage to reproduce what has been observed in the real world (of course, different domains will have different tolerances to such discrepancies).  

From a theoretical perspective, an agent-based model can be seen as both a generator and a classifier of systems. The totality of the set of systems that can possibly be generated computationally is determined by (i) the agent-based model; (ii) the updating rules (which can be seen as an extension of the model)the updating rules (which can be seen as an extension of the model); (iii) the set of parameter value combinations that can be represented, including the initial conditions and the set of possible values for random generator seeds for stochastic models (e.g. $x_{1}=[0.00000000000, 0.9999999999] \times x_{2}=[0.00000000000, 0.9999999999] \times x_{3}=[0.00000000000, 0.9999999999]$).

Correspondingly, the abstractly defined unparametised agent-based model can be seen as defining a set of systems, with subsets defined by specific combinations of (i), (ii) and (iii). Even more generally, any feature that can be represented computationally in terms of the model, either as simulation input (as in the case of (i), (ii) and (iii)) or as some property or behaviour `observed' in the simulation (see Section \ref{sec:observation} below), can be seen to define a subset of distinguishable systems and hence be used to classify simulations (see Figure \ref{fig:popSims}).

\section{`Levels' and `observations' within simulations}
\label{sec:observation}
Although agent-based models were initially motivated by the desire to understand how phenomena observed at one level can give rise to phenomena observed at another level. surprisingly little work has focused on formally defining levels or observations in agent-based simulations. This section addresses this issue by showing how to formally represent observations at different levels in agent-based modelling terms. 

\begin{figure}
	\centering
		\includegraphics[width=.9\columnwidth]{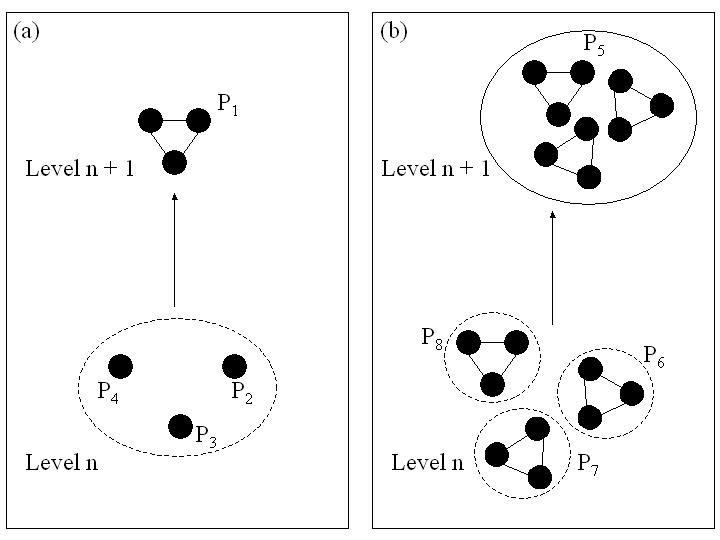}
	\caption{Two categories of hierarchy. (a) Compositional hierarchy/$\alpha$-aggregation: $P_{2}$, $P_{3}$ and $P_{4}$ are constituents of $P_{1}$. We can also say that $P_{1}$ has a greater scope than its constituents. (b) Set membership hierarchy/$\beta$-aggregation: $P_{6}$, $P_{7}$ and $P_{8}$ fall in the set defined by $P_{5}$. We can also say that $P_{5}$ has a lower resolution than its members $P_{6}$, $P_{7}$ and $P_{8}$.}
	\label{fig:alphaBeta}
\end{figure}

In order to do this, we begin by first defining what we mean in general by observing a system at different levels, and what it means to say that a property exists at a particular level. An important point to note is that the notion of level is by its very nature a relative one; it only makes sense to to say that some property exists at a higher level than some other property. Essentially there are two types of relation that link lower level properties to higher level ones:
\begin{enumerate}
	\item Composition, where lower level properties are the constituents of the higher level property in some structured relation (e.g. Na + Cl -> NaCl)\footnote{Note that structure here is meant in the most general sense here and does not necessarily imply spatial structure};
	\item Set membership, where lower level properties belong to a set defined by the higher level property (e.g. dog -> mammal). (See Figure \ref{fig:alphaBeta}.)
\end{enumerate}
In many cases, these two types of relations are combined. For example, in the case of `marriage', not only does the property require the participation of two individuals in some structured relation, but it is also blind to which particular individuals participate in this structured relation. This can be formally represented as a hypernetwork \citep{johnson06}, \citep{johnson07} or `heterarchy'  \citep{gunji04}. Furthermore, when speaking of levels, it is impossible to separate a property's existence at a particular level from the observation or description of the property at this level. The resolution or precision of observation is equivalent to set membership (since a lower resolution implies more members belonging to the set), while the scope of observation is related to composition (since a greater scope implies more constituents) \citep{ryan07}, \citep{prokopenko09}.

\subsection{Static and dynamic properties in simulations}
\label{subsec:properties}
Properties in agent-based simulations can be either static or dynamic. In terms of computational representation, static properties are subsystem states, which are represented by the values of a subset of the variables (which might also cut across agent boundaries, as in the example of group states, which take an aggregate of only a subset of the variable values within each agent member). Dynamic properties (or behaviours) are represented computationally by (possibly temporally extended) structures of state transition rule executions and state transitions. Indeed, every distinguishable system generated by an agent-based model can be described formally as a unique structure of $STR$ executions and their state transitions. 

\subsubsection{Static properties as variable values and their configurations}
At any given point in time during the simulation, we can formally describe the current state of the system as a structured set of variable values. Furthermore, we can give descriptions of this structured set at different levels. For example, from a single-agent level, the current state is described simply as the set of state variable values encapsulated in the agent. On the other hand, we can give descriptions that cut across agent boundaries, for example taking only a subset of different agents' variable values (returning to the example of a marriage, we do not necessarily need to know the colour of agents' hair to obtain the number of married couples in the system at a given time, only the marital status).

To capture the observations or description of properties, we introduce the notion of \textit{types}. A type is a specification for a class of objects such that objects satisfying the specification belong to the set defined by the class. To formalise the observation of properties in simulation using the two notions of hierarchy (compositional and subset, as defined above), we define a subsystem state type ($SST$) using a hypergraph representation where the hyperedges can be either compositional or set relations (as defined by above). A hypergraph is a generalisation of a graph, where instead of the edges being limited to binary relations between two nodes, they can be n-ary between any number of nodes. An $SST$ is then recursively defined by the hypergraph: 
\begin{equation}
SST::(\{SST\}, \{R\})|(VAR, [RG]),
\end{equation}
where:
\begin{itemize}
	\item $R$ is a compositional or subset relation connecting $n$ $SST$s 
	\item $VAR$ is a variable;
	\item $[RG]$ is the range of values that the variable must fall within (to represent the property).
	\item ($|$ stands for $OR$)
\end{itemize}
So for example, to observe marriage, we might define the $SST$:
\[
sst_{Marriage} = (\{(sst_{M1}), sst_{M2}, SST_{M3}, sst_{M4}\}, \{(sst_{M1} \wedge sst_{M2} \wedge sst_{M3} \wedge sst_{M4})\}),
\]
where
\begin{itemize}
\item $sst_{M1}=(husbID, NotNull)$ (an agent has a husband);
\item $sst_{M2}=(wifeID, NotNull)$ (an agent has a wife);
\item $sst_{M3}=(agentID, husbID)$ (identifies which agent the husband is);
\item $sst_{M4}=(agentID, wifeID)$ (identifies which agent the wife is);
\item $wedge$ stands for AND and is a compositional relationship. 
\end{itemize}

\subsubsection{Behaviours as events and structured event executions}

Given that an important motivation for agent-based modelling is often to better understand the relationship between micro-level mechanisms (represented by $STR$s) and higher level phenomena, we further distinguish between behaviours arising from the execution of a single $STR$ and those arising from an execution structure of $STR$s. In general, a structure of $STR$ executions and their state transitions is called a complex event. When a state transition results from only a single $STR$ execution, we call it a \textit{simple event} (a simple event is also a complex event, albeit one which results from only one $STR$ execution). Each simulation is therefore a complex event.

As with states, \textit{observation} of behaviour is formally represented using event types, where an event type is a specification defining a set of events (state transitions). To respect the distinction between events arising from the execution of a single $STR$ and those arising from more than one $STR$ execution, simple event types ($SET$s) are those event classes where the requirement for class membership is determined at least in part by which $STR$ is executed. However, for a given $STR$ execution, different observations (descriptions) are possible. For example, an $str_{i}$ that results in the state transition $(var1, var2) \rightarrow (var1', var2')$ can be described with three distinct $SET$s (or observed at three different `levels'):
\begin{enumerate}
	\item $\{str_{i}: [(var1, var2) \rightarrow (var1', var2')]\}$;
	\item $\{str_{i}: [var1\rightarrow var1']\}$;
	\item $\{str_{i}: [var2 \rightarrow var2']\}$;
\end{enumerate}

Furthermore, executions of different $STR$s might give rise to different state transitions, but should be defined by distinct $SET$s (i.e. $str_{i}: [var1 \rightarrow var1'] \neq STR_{j}: [var1 \rightarrow var1']$). 

Formally therefore, an $SET$ is defined both by a set of two-tuple:

\begin{equation}
SET::(STR, ST),
\end{equation}
where 
\begin{itemize}
\item $STR$ is a state transition rule, and
\item $ST::SST \rightarrow SST'$ is a constraint that the description (or observation) of the resulting state transition $SST \rightarrow SST'$ must satisfy. \footnote{In the above example, we can express $(var1, var2) \rightarrow (var1', var2')$ in $SST$ terms as $sst_{A}\rightarrow sst_{B}$, where $sst_{A}=(\{(var1,rg1),  (var2, rg2)\}, \{AND\})$ and $sst_{B}=(\{(var1,rg1'), (var2, rg2')\}, \{AND\})$ ($rg1$ and $rg1'$ represent different value ranges for var1; $rg2$ and $rg2'$ represent different value ranges for $var2$)}
\end{itemize}

So, for example, the $SET$ $\{STR_{i}: [var1 \rightarrow var1'], STR_{j}: [var2 \rightarrow var2'], STR_{k}: [var3 \rightarrow var3']\}$ would be the set of events resulting from either $STR_{i}$, $STR_{j}$ or $STR_{k}$ observed at the one variable level which satisfy the constraints satisfied (e.g. $var1>x, var1'<y...)$.

Complex event types ($CET$s) are event classes defined by a structure or set of structures of state transitions resulting from a set of structured $STR$ executions (this would include $SET$s, since $SET$s are simply classes of events where the structure of $STR$ execution is a single execution). As with $SST$s, this can be defined as a hypergraph of $CET$s, where each hyperedge can be either a compositional (structural) or subtype (set membership) relation. As in the case of $SST$s, we are thus able to integrate the two types of hierarchy (compositional and set) introduced above within a common event type. The formal recursive definition can be given as:

\begin{equation}
CET::(\{CET\}, \{R\})|SET,
\end{equation}
where:
\begin{itemize}
	\item $R$ is a compositional or subset relation connecting $n$ $CET$s 
	\item ($|$ stands for $OR$)
\end{itemize}

This definition is mainly for formal purposes. While it is possible to specify a $CET$ explicitly by defining the relationships between its constituent or subtypes, this is not always possible in practice since these relationships are not always known or, if they are, it would be extremely cumbersome to specify them in the representation above. Indeed, the goal of simulation may be to discover such relationships. In practice therefore, it is more feasible to specify $CET$s implicitly using aggregated state variables; for example, we might specify a $CET$ that includes all those structured events where a change in systemic variable $X$ (e.g. mean population crime rate) exceeds a given threshold $a$. One could then discover the execution structures after simulation by examining the simulations where $X$ exceeds $a$. 

Table \ref{tab:CETSET} outlines the empirical equivalents of the $SST/CET$ constructs defined above and gives examples of empirical data to which they can be mapped.

\section{Inter- and multi-level validation of agent-based models with empirical data}

Having defined how we can `observe' the dynamic instantiation properties and behaviours in simulation, we can also use these to classify the set of systems generated by an agent-based model (just as we can use input parameter configurations to classify systems). The repertoire of models that we can study has therefore been extended from hypotheses about how agent-level rules generate systemic properties, to hypotheses about how agent-level rules generate \textit{relationships between} systemic properties. 

\subsection{Inter-level models and validation}
Graph-based representations such as structural equation models and Bayesian networks have been used in the social sciences to describe structures of related phenomena (usually represented as variables) and the nature of the relationships (e.g. their strength, positivity). We call these \textit{structural models}. Combined with the $SST$/$CET$ framework defined above, we have a means to represent structured, defined relationships between phenomena at different levels \textit{in terms of the agent-based model itself}, and not only as ad-hoc system-level state variables. We call these \textit{inter-level models}.  

\begin{figure}
	\centering
		\includegraphics[width=.9\columnwidth]{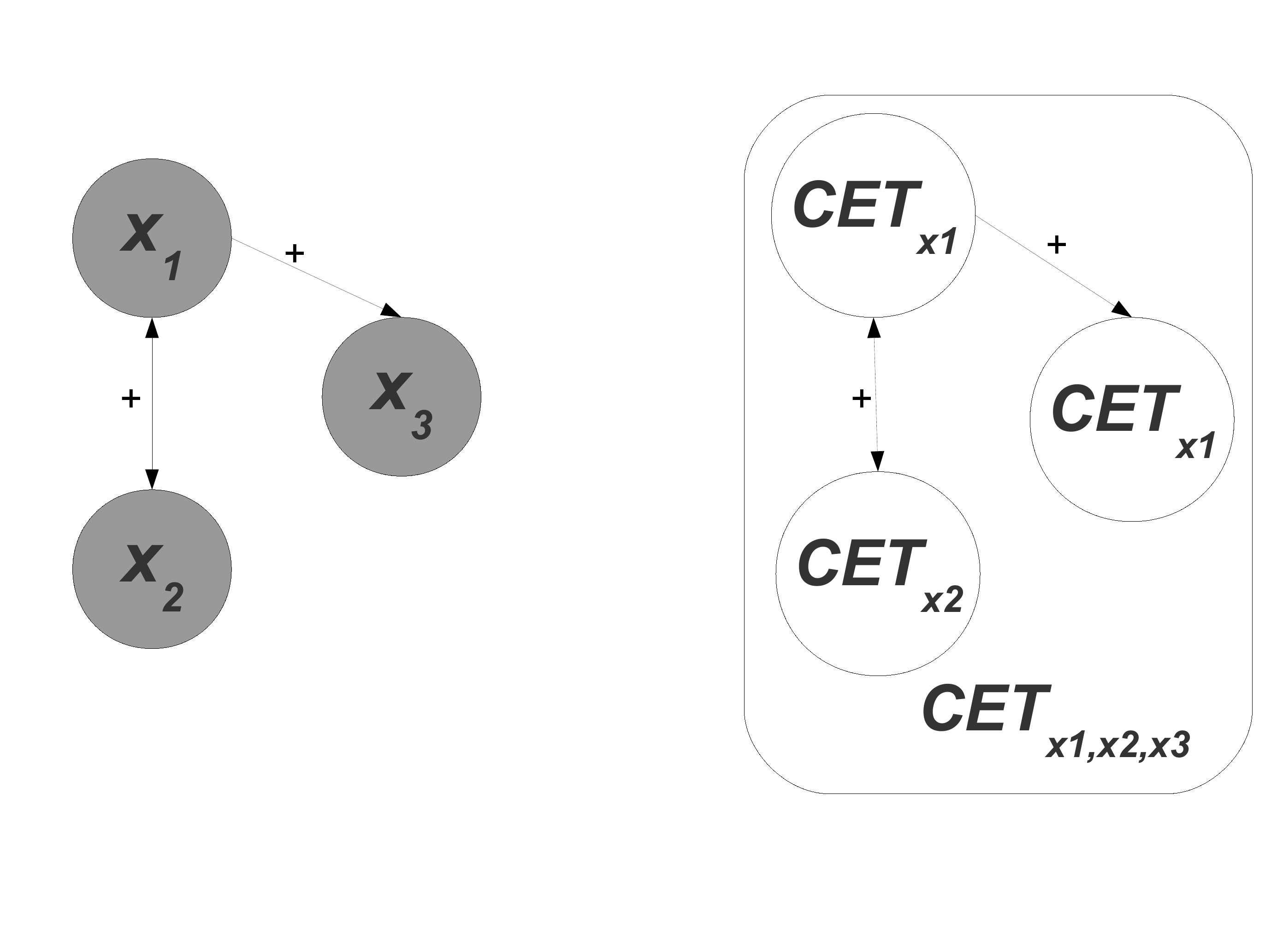}
	\caption{Left: Example of an inter-level model where the nodes in the graph, $x_{1}$, $x_{2}$ and $x_{3}$ can represent phenomena at different levels. The edges between the nodes represent  statistical associations between $x_{1}$, $x_{2}$ and $x_{3}$. These can be heterogeneous in terms of their nature (correlation, modular, causal), direction, and strength. 
Right: Formally, the inter-level model is an implicit specification for a $CET$ since the statistical associations between phenomena at different levels define the relative value ranges that must hold for $x_{1}$, $x_{2}$ and $x_{3}$ (which in turn specify further $CET$s).}
	\label{fig:interLevelModelEG}
\end{figure}

To give a concrete example, as illustrated in Figure \ref{fig:interLevelModelEG}, if variables $x_{1}$, $x_{2}$, and $x_{3}$ respectively (i) overall crime rate, (ii) clan marriage rate, and (iii) clan size, we can ask whether the agent-based model is able to generate an inter-level model such that $x_{2}$ is positively associated with $x_{1}$, and $x_{1}$ increases $x_{3}$ (Value ranges for $x_{1}$, $x_{2}$, and $x_{3}$ are also implicit specifications for three different $CET$s). Assuming that the agent-based model was developed and parametised in an empirically-driven fashion, we would require multiple data sets with data corresponding to $x_{1}$, $x_{2}$ and $x_{3}$ to validate the inter-level model. If the associations specified by the model are found in these empirical data, the inter-level model is said to be valid, in the sense that it has not been shown to be wrong. \footnote{The precise type of association relationship e.g. correlation, mechanistic causation, phenomenal causation, depends on the statistical constraints that need to be satisfied; these would depend on the goals of the modelling project.}.

Similarly, if we have data corresponding to variations in parameter values (e.g. different policies at the individual level, which could be translated into agent propensities for action), we can hypothesise about the effects of interventions at the agent level on the structural or inter-level model. Or, if we have very little information about what might be going on at the individual level, we can classify simulations into those which generate these inter-level relationships and those which do not (or do so with a far weaker degree), and then conduct further analyses to determine what the `unsuccessful' simulations have in common. This might involve specifying and identifying further $CET$s or, more simply, evaluating $SET$ frequencies (and hence agent level $STR$ execution frequencies). If, say, we find that a given $SET$ is associated with an inter-level model, it would be worth probing further on the effects of the particular $STR$ associated with this $SET$. In real world terms, this might, for example, correspond to identifying a particular law as being associated with a self-perpetuating web of social problems.

\subsection{Multi-level models and parameter spaces}

Given that an agent-based model aims to represent the essential individual-level mechanisms underlying systemic phenomena, a deeper understanding of these mechanisms can only be attained through probing the model's behaviour under different conditions. In practice, this is done through systemtically varying the model's parameters, which (either individually or together) can be used to represent different real-world scenarios. A characterisation of the parameter space can therefore be seen as a statement of how our modelled mechanisms interact under different conditions.

The multi-level statistical framework has proved to be extremely promising in the analysis of data in the social sciences. In multi-level modelling (also known as hierarchical linear models), effects can vary depending on the level of analysis. For example, a model relating two variables $q$ and $s$, representing say, the salary per year of an individual and an individual's level of education, and a parameter $p$, representing age, we might find that different levels (precisions) of $p$ grouping expose different relationships or relationship strengths. If we choose a precision of 1 year to group individuals (i.e. 1, 2, 3....), there may be little difference between groups, while a precision of 10 years (i.e. 1-10, 11-20, 21-30...) might yield a stronger relationship between $q$ and $s$ for some groups than for others. 

This same framework can be applied to the parametisation of agent-based models. If, say, an agent-based model has two parameters $p1$ and $p2$, we can probe the model by simulating with different $p1 \times p2$ configurations, giving an $n_{1} \times n_{2}$ matrix of $CET$s, with each matrix cell corresponding to simulation under the particular $p1 \times p2$ configuration ($n_{1}$ is the set of values for $p1$ we simulate with, and $n_{2}$ is the set of values for $p2$).

\begin{figure}
\label{fig:parameterSpace}
\begin{center}
\begin{tabular}{c}
\includegraphics[width=.6\columnwidth]{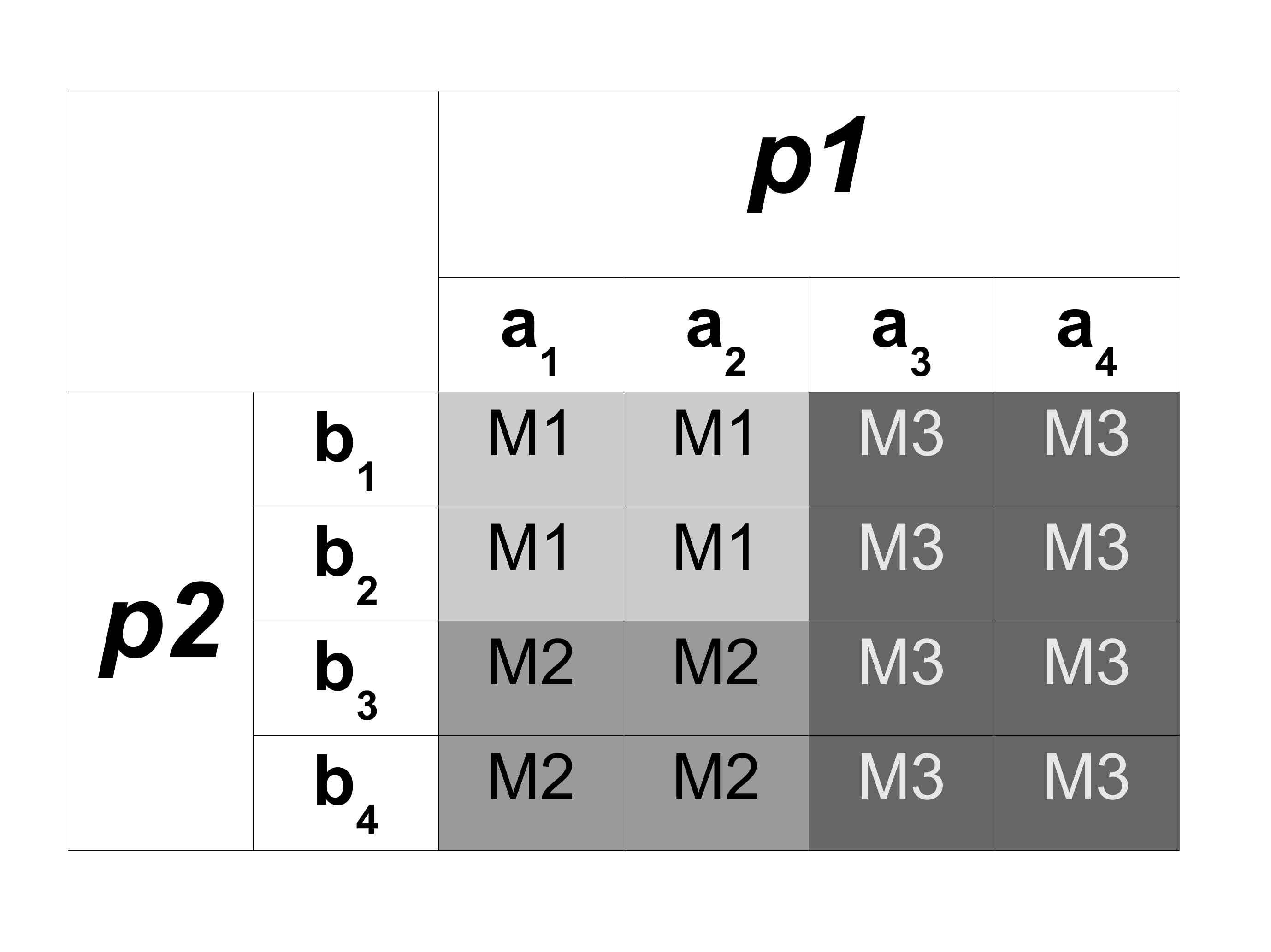}\\
\includegraphics[width=.6\columnwidth]{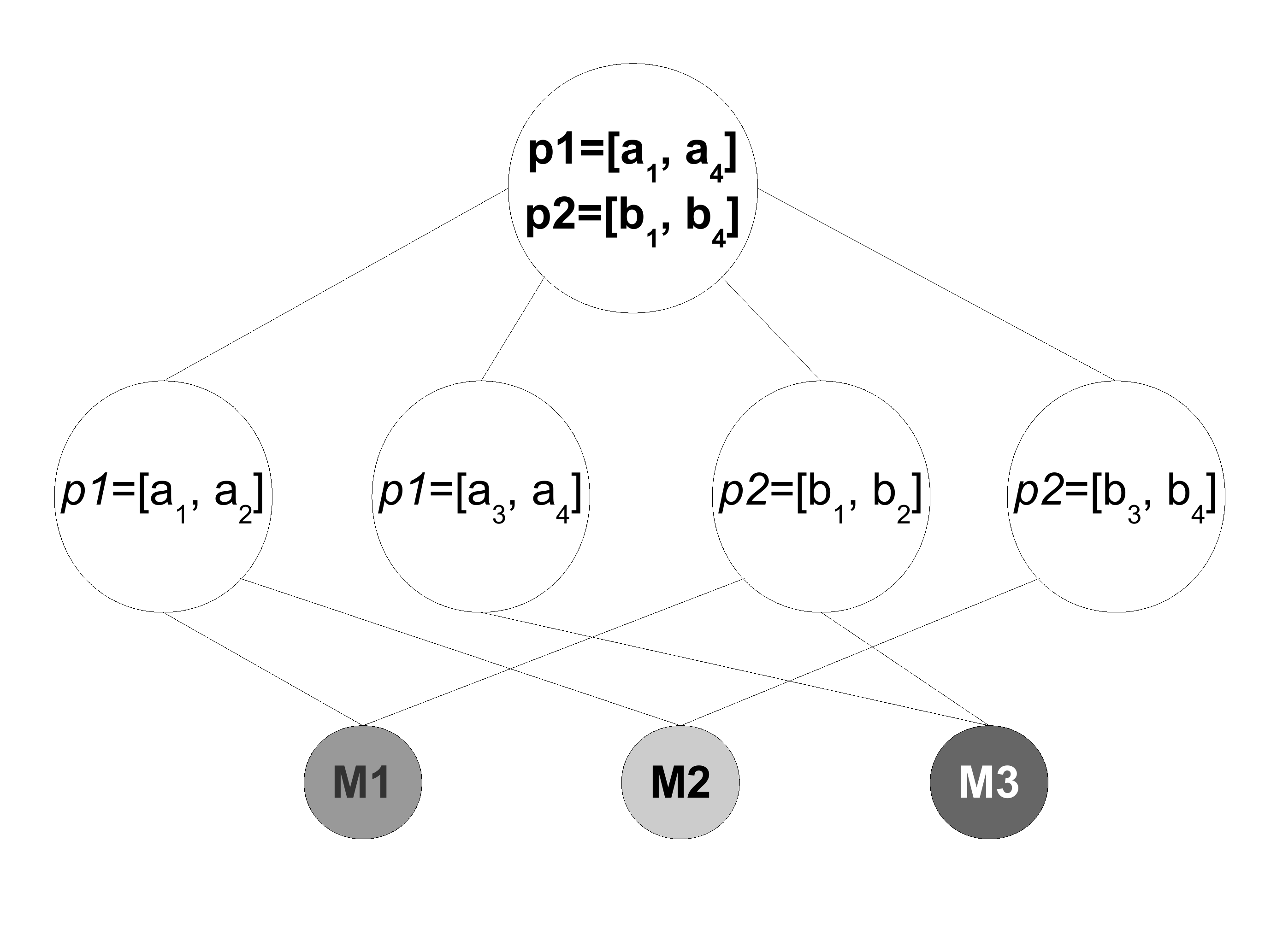}
\end{tabular}
\end{center}
\caption{Top: Matrix representing different models (system behaviours) for different parameter ranges of $p1$ and $p2$. $M_{1}$, $M_{2}$ and $M_{3}$ might  be radically different models (e.g. $M_{1}$ might represent a simple linear relation while $M_{2}$ could be an inter-level network relation, or they could simply be different strengths of of the same model structure.
Bottom: Multiple multi-level models represented in a single hierarchy (`heterarchy'), where each node also represents a distinct $CET$.
Within the range $p1=[a_{1}, a_{2}$, $M1$ and $M2$ can be treated as submodels defined by two different $p2$ value ranges: $[b_{1}, b_{2}]$ and $[b_{3}, b_{4}]$. Within the range $p2=[b_{1}, b_{2}]$, there are also two submodels, $M1$ and $M3$. Hence, $M1$ can be multiple classified as a submodel of both $p1=[a_{1}, a_{2}$ and $p2=[b_{1}, b_{2}]$.
All three models, $M1$, $M2$, and $M3$ can be treated as submodels of the multi-level model defined by the range $p1=[a_{1}, a_{4}]$ and $p2=[b_{1}, b_{4}]$.}
\end{figure}

If some region of this matrix contains $CET$s differing greatly from the rest of the matrix (but similar to each other), we separate it from the remainder of the matrix using the $p1$ and $p2$ values. For example, we could discover a multi-level model in which $M_{1}$ holds between ranges $p1=[a_{1},a_{2}]$ and $p2=[b_{1}, b_{2}]$; $M_{2}$ holds between  ranges $p1=[a_{1},a_{2}]$ and $p2=[b_{3}, b_{4}]$; and $M_{3}$ holds between  ranges $p1=[a_{3},a_{4}]$ and $p2=[b_{1}, b_{4}]$, where $M_{1}$, $M_{2}$ and $M_{3}$ could be any specified relationship, from simple linear correlation to an inter-level network model. (In terms of $CET$s we can also say that the $CET$ associated with $M1$ and the $CET$ associated with $M2$ are both subtypes of a third $CET$ defined by the parameter range $p1=[a_{1}, a_{2}]$.) Figure \ref{fig:parameterSpace} illustrates this. As in the case of inter-level models, the multi-level model itself implicitly specifies a $CET$, as do its sub-models.

Regions in which parameters (either on their own or in combination) are particularly sensitive are regions in which the resolution defining groups has to be higher to observe the differences. Complexity comes in when the levels are defined irregularly (i.e. the resolution for defining groups varies; this can be within or between dimensions). To validate these regions, we need to also split the empirical data into the appropriate groupings, possibly requiring relatively high resolution data for some ranges. 

In the above example, we would need two empirical datasets corresponding to the two interval $p2=[b_{1}, b_{2}]$ and $p2=[b_{3}, b_{4}]$ within $p1=[a_{1}, a_{2}]$. These two datasets correspond to the two groups (`levels'): $(p1=[a_{1}, a_{2}] \times p2=[b_{1}, b_{2}])$ and $(p1=[a_{1}, a_{2}] \times p2=[b_{3}, b{4}])$. A third dataset is required for the group $(p1=[a_{3}, a_{4}] \times p2=[b_{1}, b_{4}])$. If, in these data groupings, the relationships defined by $M_{1}$, $M_{2}$ and $M_{3}$ hold, the multi-level model generated through simulations can be said to have been validated by the empirical data. \footnote{Of course, when we wish to establish stricter, more specific relationships between models and parameters (e.g. causal relationships), validation becomes more problematic, since it is then necessary not only to show the same irregular regions show up in empirical data as in simulation-generated data, but also that they do so for the correct reasons. For example, should we say that $p1$ \textit{must lie within} $[a_{1},a_{2}]$ for $M1$ and $M2$ to hold for $p2=[b_{1}, b_{2}]$ and $p2=[b_{3}, b_{4}]$, or is it that in the value range $[a_{1}, a_{2}]$, $p1$ \textit{has no effect} when $p2$ lies between $b_{1}$ and $b_{4}$? The difficulty of validating such relations is a general one however, and the challenge comes mainly from finding the appropriate `treated' and untreated' cases. This can be particularly challenging in the social sciences, since assumptions often have to be made about the commonalities between two cases since active treatment (the methodology of the experimental sciences) is not usually appropriate (one could even argue that it is inconsistent with the very point of the social sciences). Data that would allow us to distinguish, for example, necessary conditions from irrelevant background conditions, are therefore extremely difficult to obtain.}   

This multi-level approach to describing the state space of an agent-based model maps more naturally to data obtained from empirical studies than the equation-based descriptions of phase transitions typically used to characterise complex systems by physicists while still being formally related to this description.


\section{Summary and conclusions}
In this article, we have introduced subsystem types ($SST$s) and complex event types ($CET$s), which allow us to formally describe or `observe' at any level of abstraction the states and behaviours generated by an agent-based model. Therefore, we can characterise an agent-based model as a function that generates a set of $SST$s and $CET$s with given probability distributions.\footnote{However, if an agent-based model contains real values or stochasticity, the computational representation of the model will only be an approximation, and the set of computationally generated $CET$s (simulations) generated may be a biased sample from the true set of systems that could be generated by the model.}

$SST$s and $CET$s can be used as the building blocks for defining sophisticated inter-level and multi-level models (formally speaking, inter-level models and multi-level models are also $CET$s). Structural and inter-level models allow us to define a structure of statistically related $CET$s and/or $SST$s, and the types of statistical relationships that need to hold between them. The multi-level modelling framework allows us to define different classes of system for which different models hold (models might be structural, inter-level, or simple linear models). This can also be linked to the sensitivity of parameters and the characterisation of the model's parameter phase space. 

From a more practical perspective, the ability to specify structured statistical relationships between phenomena at different abstraction levels in ABM terms allows us to formally define the isomorphism between models and empirical observations and data. Networks and hierarchies of statistical associations then give us more stringent sets of criteria for empirically validating these types of models. Rather than simply requiring that an agent-based model can generate phenomenon $X$ for example, we can stipulate that it should be able to generate associations with particular strengths between phenomena $X$, $Y$ and $Z$ in scenario $A$, and a different set of strengths in scenario $B$. By identifying emergent structures of behaviour, we are able to formally relate the agent-based model to empirical observables. This represents a significant step towards true integration of empirical and model-driven research in the social sciences. 

\newpage

\begingroup

\parindent 0pt

\parskip 2ex

\def\enotesize{\normalsize}

\theendnotes

\endgroup

\newpage

\bibliographystyle{asa}
\bibliography{ABMAnalysis}
\newpage
\section{Figures and tables}

\newpage

\newpage

\newpage

\newpage

\newpage

\newpage

\newpage

\newpage

\newpage

\begin{table*}
	
		\begin{tabular}{|p{3cm}|p{6cm}|p{6cm}|}
		\hline
		&&\\
		\textbf{} & \textbf{Empirical equivalent} & \textbf{Validation data}\\
		&&\\
		\hline
		&&\\
		$SST$ & Observed situation in a system at a given point in time & Individual, collective, population measures and/or statistics e.g. an individual's current employment status, an organisation's current revenue, a country's GDP at time $t_{i}$ \\
		&&\\
		\hline
		&&\\
				$STR$ & Hypothesised micro-level (which can be individuals, organisations, countries depending on what the agents are modelling) responses to environment. e.g. if individual unable to pay bills and feels cheated, more likely to steal; if tax imposed on activity $A$, firm less likely to do $A$. & May be largely theory-based, so data not always available. If available, may be from experimental or case studies at micro-level e.g. Social Psychology studies investigating the responses of human subjects, case studies on firms. \\
				&&\\
		\hline
		&&\\
		$SET$ & Micro-level behaviour in a system that arises as a direct consequence of the entity's response to his/her/its environment e.g. stealing when unable to pay bills. & Data from social/behavioural/cognitive psychology studies and/or case studies (especially when the entity is an organisation or geographical region). \\
		&&\\
		\hline 
		&&\\
		$CET$ (includes $SET$s)& Observed behaviour in a system. As well as micro-level behaviour, this also includes collective or systemic behaviours at other levels e.g. increase in criminal activity in community X. & Data from experimental studies and/or case studies addressing micro-level behaviour; population statistics and changes in population statistics over time.\\
		&&\\
		\hline 
		
			\end{tabular}
	\caption{Table outlining the empirical equivalents and validation data for different constructs in the $SST/CET$ framework.}
	\label{tab:CETSET}
\end{table*}

\end{document}